\documentstyle[11pt,newpasp,twoside,epsf]{article}
\def\edcomment#1{\iffalse\marginpar{\raggedright\sl#1\/}\else\relax\fi}
\reversemarginpar

\begin{document}
\title{FeII/MgII, [Fe/Mg] Ratios and High-z Quasars}
\author{Kirk Korista and Nalaka Kodituwakku}
\affil{Western Michigan University, Department of Physics, Kalamazoo,
MI 49008-5252}
\author{Michael Corbin}
\affil{Space Telescope Science Institute, 3700 San Martin Drive, Baltimore,
MD 21218}
\author{Wolfram Freudling}
\affil{Space Telescope European Coordinating Facility,
Karl-Schwarzschild-Strasse 2, 85748 Garching, Germany}

\begin{abstract}
It has been suggested in the literature that the (Fe/$\alpha$) abundance
ratio may be used as a chronometer, due to a delay in this ratio reaching
its solar value as predicted by galactic chemical evolution models.
Using grids of photoionization models along a sequence of the (Fe/Mg)
abundance ratio vs.\ metallicity with time in a giant elliptical starburst
scenario, we investigate the relationship between the (Fe/Mg) abundance
ratio and the FeII/MgII emission line flux ratio under the assumption
that these lines originate in photoionized clouds within the broad
emission line regions of quasars.
\end{abstract}

\vspace{-0.5cm}

\section{Introduction}

Nearly 20 years ago, in their paper investigating the physics of
FeII emission in quasars, Will, Netzer, \& Wills (1985; WNW) suggested
the importance of measuring the (Fe/Mg) abundance ratio in quasars, in
that an overabundance of iron in quasars  and active galaxies ``may have
important implications for their evolution.'' Hamann \& Ferland (1993;
1999) discussed the possible use of the FeII/MgII $\lambda$2800 emission
line flux ratio as a chronometer in quasars. With Type~II SNe producing
most of the $\alpha$ elements (O, Mg, etc), and Type~Ia SNe accounting
for most of the iron enrichment, a characteristic delay of $\sim$~1 Gyr
is expected for (Fe/$\alpha$) to reach solar values. However, the value
of this delay has been a matter of recent controversy, with some models
of Matteucci \& Recchi (2001) suggesting much shorter delay time scales,
depending upon the environment.

Recent quasar surveys, such as the Sloan Digital Sky Survey, have
discovered large numbers of high redshift quasars. Infrared spectroscopy
of these quasars have revealed their emission in MgII and UV-optical
FeII. Elston et~al.\ (1994) was the first to report strong FeII emission
in $z > 3$ quasars. Since then several investigators have measured the
UV FeII/MgII emission ratio in quasars with redshifts up to $z \approx
6$, with hopes of gauging early chemical evolution in the centers of
host galaxies and pinning lower limits to the age of the universe at
high redshift (Kawara et~al.\ 1996; Thompson et~al.\ 1999; Aoki et~al\
2002; Dietrich et~al.\ 2002; Iwamuro et~al.\ 2002; Pentericci et~al.\
2002; Freudling et~al.\ 2003; Dietrich et~al.\ 2003a; Maiolino et~al.\
2003; Barth et~al.\ 2003). Yoshii et~al.\ (1998) used galactic chemical
evolution models and FeII/MgII measurements of a $z = 3.62$ quasar to
constrain cosmological models, based upon the delay in the iron enrichment
and two important assumptions. First, as suggested by WNW, they assumed
that a flux ratio FeII(UV $+$ opt)/MgII $\approx 3$ represents solar
(Fe/Mg). This has never been established theoretically. Second, they
assumed that the flux ratio scales directly with the abundance ratio. It
is this latter relationship that we investigate, as an important first
step toward possibly establishing the FeII/MgII flux ratio in quasars
as a meter stick of time and/or (Fe/$\alpha$) enrichment history (see
Di~Matteo et~al.\ 2003). A linear relation (logarithmic slope $\approx
1$) between the flux and abundance ratios is not expected, due to the
thermostatic effects of significant coolants. The logarithmic slope in
the relation may even change since MgII is only a doublet, while FeII
can redistribute its cooling to a vast array of UV and optical multiplets
(where one may not be measuring).

\section{Grids of Photoionization Models}

Using Gary Ferland's photoionization code, Cloudy, we computed 8 separate
grids of photoionized clouds, each grid containing 841 models and spanning
7 decades in hydrogen number density ($7 \leq \log~n_H(\rm{cm^{-3}})
\leq 14$) and ionizing photon flux ($17 \leq \log~\Phi_H(\rm{cm^{-2}
s^{-1}}) \leq 24$) for a fixed cloud total hydrogen column density
($10^{23}$~cm$^{-2}$). Each grid was assigned elemental abundances and
metallicity ($Z$) from the giant elliptical starburst abundance set of
Hamann \& Ferland (1999; their Figure~13) for a particular age of the
starburst (see Figure~1). These spectral simulations used the default
simplified model FeII atom within Cloudy. The predicted quasar emission
line spectrum from each grid was determined by summing the emission
over a broad distribution of clouds within the grid following the
``locally optimally-emitting clouds'' (LOC) scheme of Baldwin et~al.\
(1995). In addition Gary Ferland and Jack Baldwin have kindly provided
predicted FeII(2240-2660~\AA\/)/MgII ratios for 3 simple combinations of
$Z$ and (Fe/Mg) (Figure~1; solar, high \& low $Z$), each of these grids
computed using Cloudy's detailed {\em 371-level} model FeII atom. The
spectral region 2240-2660~\AA\/ in FeII emission will be referred to as
the FeII(UV bump).

\begin{figure}
\plotfiddle{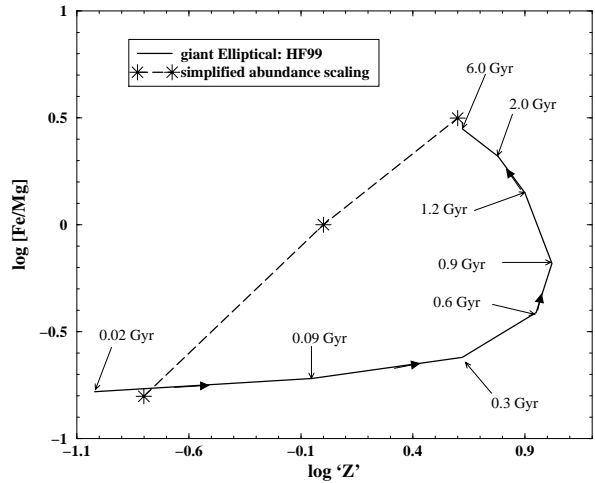}{2.3in}{-90}{40}{40}{-145}{215}
\caption{The time-dependent giant elliptical starburst (Fe/Mg) abundance
ratio vs.\ metallicity (relative to solar), as adopted from Hamann \&
Ferland (1999) and utilized in the present computations (at each of the
8 marked starburst ages). Solar metallicity is reached well before solar
(Fe/Mg); note also the rapid rise in the latter near a starburst age of
$\sim$0.5 Gyr. Also shown are three simplified abundance sets representing
low, solar, and high $Z$.}
\end{figure}

\section{Results}

For each of our 8 grids, the total cooling from the simplified FeII model
atom was integrated over a broad distribution of clouds, as described
above, and ratioed to the emission from MgII $\lambda$2800. The same was
also done for the 3 grids that utilized the detailed model FeII atom,
except that the predicted emission from the more commonly measured FeII(UV
bump) was used. The flux ratios from the 8 grids utilizing the simplified
FeII model atom were normalized to the FeII(UV bump)/MgII flux ratio
from the solar metallicity grid predicted by the detailed FeII model
atom. The slope in the $\log $(FeII(UV bump)/MgII) vs.\ $\log $(Fe/Mg)
relation from our 8 grids was found to be 0.4, whereas that between the
3 grids utilizing the detailed model FeII atom was found to be 0.6. As
expected the predicted logarithmic slope is substantially less than 1.

In Figure~2 we show the FeII(UV bump)/MgII flux ratio as a function the
approximate age of the starburst. The flux ratio climbs rapidly and then
flattens beyond $\sim$1.5 Gyr, corresponding to a (Fe/Mg) abundance ratio
of about twice solar (Figure~1). Based upon these results, we might expect
to measure a systematic decline in the FeII(UV bump)/MgII flux ratio
in samples of quasars for redshifts $z \ga 3.7$ assuming a host galaxy
formation redshift $z_{form} \approx 20$ or $z \ga 2.7$ assuming $z_{form}
\approx 6$, for presently favored cosmological parameters. While this
flux ratio is difficult to measure, especially in the observed frame IR,
no such systematic decline has been detected to date. More importantly,
Baldwin et~al.\ (these proceedings) find that the origin of even the UV
FeII emission in quasars remains in question; part of this is reflected in
the weakness of the predicted flux ratio in Figure~2 even at large values
in (Fe/Mg). Without intending to sound obvious, we must first understand
the origin of the FeII emission in quasars before it can be used as either
a clock or as a means to track enrichment histories of the centers of
host galaxies. Additional clues to chemical evolution at high redshift
may come from other heavy element emission lines (e.g., NV, CIII]; see
Dietrich et~al.\ 2003b), from measurements of CO (Bertoldi et~al.\ 2003),
and even measurements of the X-ray iron K-edge (Hasinger et~al.\ 2003).


\begin{figure}[h]
\plotfiddle{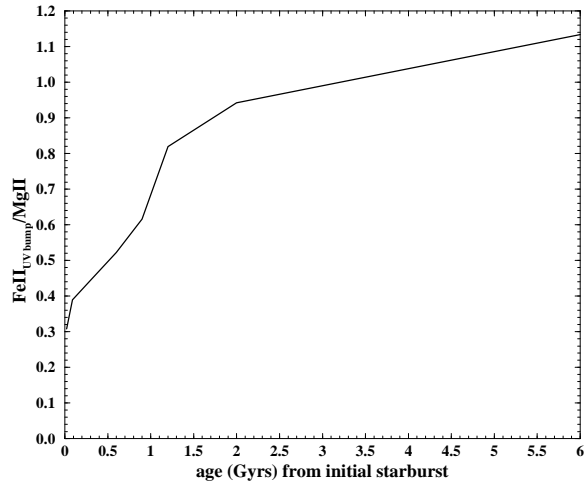}{2.3in}{-90}{40}{40}{-145}{215}
\caption{The FeII(UV bump)/MgII linear flux ratio as a function of age,
as predicted from the photoionization models using the giant elliptical
starburst abundances in Figure~1. Note the rapid change in this flux
ratio near $\sim$1.5 Gyr.}
\end{figure}


\end{document}